# ChatGraPhT: A Visual Conversation Interface for Multi-Path Reflection with Agentic LLM Support

Geoff Kimm

School of Design and Architecture, Swinburne University of Technology[1]

Linus Tan

School of Design and Architecture, Swinburne University of Technology

**Abstract:** Large Language Models (LLMs) are increasingly used in complex knowledge work, yet linear transcript interfaces limit support for reflection. Schön's Reflective Practice distinguishes between reflection-in-action (during a task) and reflection-on-action (after a task), both benefiting from non-linear, revisitable representations of dialogue. ChatGraPhT is an interactive tool that shows dialogue as a visual map, allowing users to branch and merge ideas, edit past messages, and receive guidance that prompts deeper reflection. It supports non-linear, multi-path dialogue, while two agentic LLM assistants provide moment-to-moment and higher-level guidance. Our inquiry suggests that keeping the conversation structure visible, allowing branching and merging, and suggesting patterns or ways to combine ideas deepened user reflective engagement. Contributions are: (1) the design of a node-link, agentic LLM interface for reflective dialogue, and (2) transferable design knowledge on balancing structure and AI support to sustain reflection in complex, open-ended tasks.

## 1 INTRODUCTION

The computer scientist J.C.R. Licklider wrote in 1960 that, before any possible supremacy of Artificial Intelligence (AI) over human intellect, there would be an extended period of intellectual progress driven by an "intimate association" between human and machine [28]. Since the release of ChatGPT in 2022, generative AI has emerged as a truly general-purpose technology, much like computing and electricity, whose full impact may take decades to materialise [15]. The pace of generative AI's uptake rivals the early mass-market adoption of computers in the 1980s, especially beyond work settings [7]. It is this rapid adoption and adaptability, particularly in the subdomain of large language models (LLMs), that now appears to offer precisely Licklider's intimate association in a widely accessible way.

This intimate association has aspects of a reflective partnership. Empirical studies are beginning to document how knowledge workers use LLMs as collaborators in complex cognitive work: brainstorming ideas, structuring content, and reframing problems [30,35]. These activities strongly resemble the processes described in Donald Schön's *Reflective Practice* in which practitioners learn by engaging in a "reflective conversation" with a situation [40].

Viewing this emergent practice through the lens of Schön's theory reveals a fundamental tension with mainstream LLM interfaces. First, the modality of the dominant LLM interface is linear, whereas the structure of the reflective practice it is increasingly called upon to support is inherently non-linear and recursive. Further, these interfaces lack integrated guidance on synthesis and metacognition to support a full reflective cycle. This dual challenge motivates our design inquiry.

---

[1] gkimm@swin.edu.au

This paper presents an early-stage research prototype and reports preliminary evaluative observations. It makes two main contributions to Human-Computer Interaction and generative AI interface design: it introduces ChatGraPhT, a novel interface that combines a manipulable conversational node-link structure with agentic LLM scaffolding to support reflective dialogue, and offers design knowledge on how structural and agentic features afford and constrain reflective conversation.

## 2 FOUNDATIONS FOR A REFLECTIVE DIALOGUE SYSTEM

This section draws on established theory to build a foundation for reimagining how reflective dialogue with LLMs might be supported in practice. We first introduce Schön's *Reflective Practice* and then examine two ways current interfaces fall short of supporting it.

### 2.1 Schön's Reflective Practice

Reflective practice [40] is a way for professionals to learn by thinking about their actions while working, rather than just applying fixed rules [41]. This learning happens through what Schön calls a *reflective conversation* [42], a dialogue-like process between the practitioner and their situation. Each action, such as a design move, produces a response from the situation, which can confirm, challenge, or surprise the practitioner [42]. These unexpected responses spark *reflection-in-action*, where the practitioner thinks critically in the moment and adjusts their next steps accordingly. For example, in a design situation, the practitioner might introduce an idea, see how it changes the overall pattern, and rethink their approach in response to this new understanding. In contrast, *reflection-on-action* occurs after the situation has unfolded, when the practitioner steps back to review their earlier moves, make sense of what happened, and draw lessons for the future. Through this dialogue during and after action, practitioners adapt to uncertainty and learn through doing.

While Schön's concept of reflective practice emphasises situated learning, structuring reflections can deepen its benefits [1,37]. *Structured reflection* ensures reflection occurs regularly and gives individuals space to think holistically about the different aspects of their ideas. It uses prompts and planned moments to prevent superficial responses and supports a broader, more integrated understanding of the task [36].

### 2.2 A Disjunct of Mainstream Chat Interfaces and Reflective Dialogue

The support of reflection has long been a focus in HCI [4,6]. Despite this, we see two limitations in mainstream LLM interfaces for support of reflection: their strict linear textuality and the need to move dialogue and its outcomes into external processes for deeper reflection. We describe each of those limitations here and interaction strategies that arise from them.

*2.2.1 The Constraint of Linearity on Reflection*

Mainstream LLM interfaces such as OpenAI's ChatGPT and Google's Gemini permit a basic reflective loop: a user issues a prompt (a move), inspects the output, and issues a follow-up (an adjusted move). Users can also scroll back through past turns, offering a rudimentary basis for reflection-on-action. The linear textual flow of LLMs can impede communication of complex structures and can hinder iterative thought processes, leading to verbose and disjointed exchanges [22]. For a reflective chatbot study, Park [31:135–136] observed a linear structure provided "limited space for users to explore their own ideas" and constrained their ability to take a non-linear "detour". Viewed through the lens of Schön's theory, this structural limitation insufficiently supports reflection-in-action, which depends on the freedom to make responsive moves and explore surprising outcomes as they arise. Further, it undermines reflection-on-action by obscuring a conversation's overall structure, thereby making it difficult for a user to review and make sense of earlier moves. As shown in Figure 1, multiple possible moves in a linear LLM interface collapse into a vertical transcript, forcing back and forth navigation. This linearity increases cognitive load and hinders comparison of alternatives, as users must mentally manage disparate conversational branches.

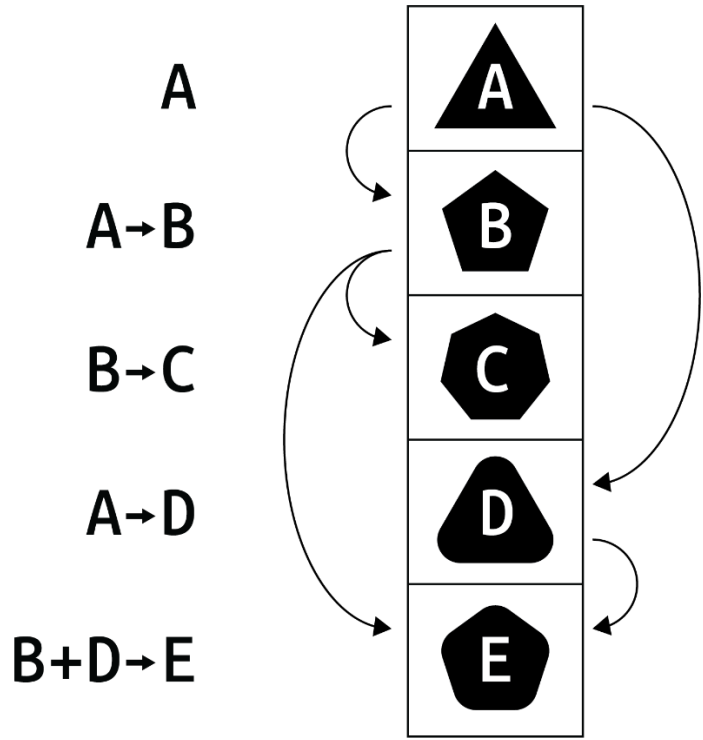

Figure 1 A to E show conceptually how several potential paths are rendered as one scrolling sequence in a typical chat interface. Each step represents a close group of user-LLM exchange.

The user-AI interplay in LLM and chatbot dialogues satisfies Herman et al.'s [20] test for visualisation by a graph: "Is there an inherent relation among the data elements to be visualized?" A graph, formally a set of vertices connected by edges [44:1], can be graphically represented in information visualisation as a node-link diagram [39]. Node-link diagrams provide the kind of visualisation of complex information spaces that can support sensemaking [43] because they make key relationships visible, group related information to reduce search, and support the easy perceptual inferences that make diagrams computationally powerful [26].

Node-link diagrams are increasingly researched for representing non-linear LLM interaction. In this role, they have been implemented as a type of directed acyclic graph [44:100] in which links have a direction of action, such as from a user prompt to an LLM response. They have been found to help overcome the limitations of strictly linear LLM interfaces [22], and support creative exploration [2,9], hypothesis testing [13,14], systems thinking [21], and information-seeking [22] by users. To our knowledge, reflection has not been the focus of node-link LLM or chatbot diagrams.

*2.2.2 A Fragmentation of Support for Reflection*

A second limitation of mainstream LLM interfaces is a lack of integrated tools to support the full arc of reflection. Effective reflection processes must often move outside the chatbot dialogue in some way. In educational contexts, external prompting or peer collaboration deepened reflection with ChatGPT when LLM-chat alone fell short [3,34]. This reliance on external processes demonstrates limited integrated support for reflection-on-action within a single, coherent environment.

An interaction strategy that responds to this limitation flows from the node-link diagram. LLMs can guide and elicit a user's conversational interactions to support deeper critical thinking and reflection [10,23,24,27,32]. Representing a conversation as a node-link diagram creates a common object, shared with the user, that an LLM can access and adjust. In such "shared representations", both parties can collectively analyse and understand a task, take action on the representation, and the AI can offer suggestions the user can accept, reject, or revise [19]. Empirical evaluation of LLM-driven graphical user interface control has shown improvements in user satisfaction and engagement [46]. The node-link precedents show aspects of shared representation control: an LLM can make suggestions a user can accept [21], can add nodes to structure domain information [22], and can insert candidates for further exploration [13,14].

Programs operating on a shared representation can show bounded agency, in the sense of an agent as a system that perceives its environment and, over time, acts to pursue its objectives [16,38]. For agentic LLMs, these behaviours can take place in a reasoning-acting-interacting cycle [33], and can include planning on a task with human feedback, memory with both natural language and structured formats, and action goals of task completion or communication with other agents or humans [45]. LLM agents, therefore, potentially offer support for user dialogue with reflective node-link diagrams.

## 3 STUDY DESIGN

This work is intended as an exploratory systems paper documenting design space, rather than an evaluative study. The study adopts a constructive design research (CDR) methodology in which the construction of design artefacts is the central means of enquiry and knowledge generation [25:1]. Unlike design research, which often centres on user studies [see 11], or Research through Design, which emphasises exploration through practice [see 17], CDR is grounded in theory-informed making and the designed artifact is both the outcome and the method of research.

Two research objectives guide the project:
1. To develop and demonstrate an interactive system that addresses the limitations of linear LLM interfaces for reflective practice by combining a manipulable, non-linear graph with agentic scaffolding.
2. To investigate how this system affords and constrains the reflective modes of reflection-in-action and reflection-on-action.

We iterated three prototypes addressing linearity and lack of metacognitive support, with Schön's concepts of reflection-in-action and reflection-on-action informing the design criteria:
- ChatGPTree: conversational branching for reflection-in-action.
- ChatBraid: node-link representation for retrospective navigation and reflection-on-action.
- ChatGraPhT: multi-conversation canvas, branch merging, and an agentic framework within a shared representation.

The prototypes were built using a web-based framework (HTML/CSS/JavaScript) and accessed the OpenAI LLM API. Evaluation followed a first-person method [12,29] combining technical experimentation with interpretive reflection. This method aligned with the study's exploratory goals, enabling an interpretive, expert-driven analysis of a novel system to generate initial findings. Each author engaged with the prototypes through the lens of a primary expertise: Author A in design computing application and theory, and Author B in reflective practice and design theory. Insights from these sessions were synthesised in collaborative reflective discussion to inform the next design iteration.

## 4 A THREE-STAGE CONSTRUCTIVE DESIGN INQUIRY

We report a three-stage design enquiry. The first two prototypes demonstrate reflective conversation support through structural changes of the typical chatbot interface. The third prototype extends those and shows how reflection can be actively scaffolded through integrated agentic support.

### 4.1 ChatGPTree: Enabling Reflection-in-Action with Conversational Branching

Our first prototype, ChatGPTree (Figure 2), supports reflection-in-action beyond a single conversational thread. ChatGPTree's interface extends the vertical, chronologically ordered list of LLM-user interactions by allowing a user to treat the conversation as a branching tree of reflective lines (Figure 2, #1 and #2). Any LLM response within the dialogue can be extended with new prompts, creating a new conversational branch if that action occurs within the interior of the conversation tree. A user could therefore return to and extend any past lines of reflection, rather than being forced to start anew or to manually reference past dialogue and fold an extension into a linear dialogue. Each branch is a distinct conversational context: the LLM receives only the dialogue history from a new user prompt back to the original conversational root (Figure 2, #3 and #4). In this example, a user can now initiate two instances of reflect-in-action in response to a single chatbot message.

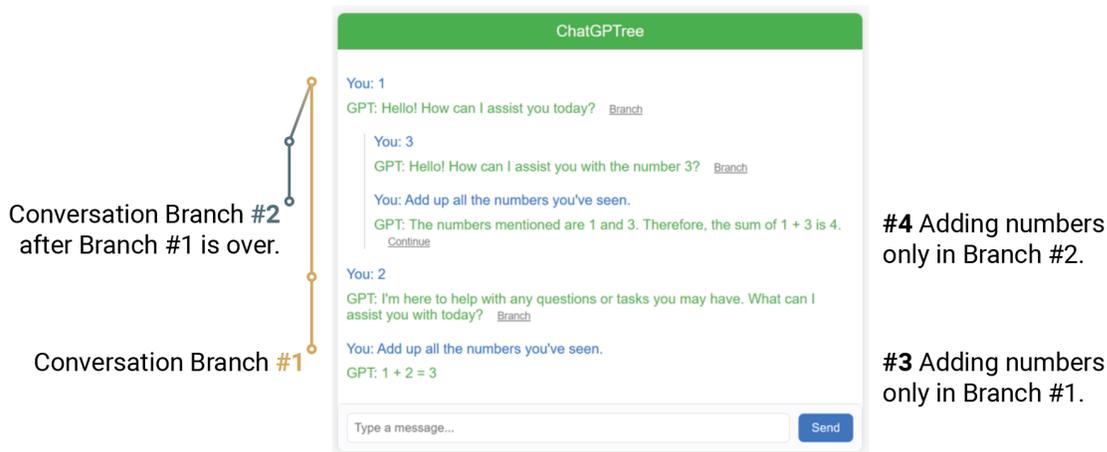

Figure 2 The ChatGPTree interface. A simple addition exercise shows branches have distinct dialogue histories.

We investigated the prototype's utility through an author-based inquiry with a reflective exercise (see Figure 3). This process revealed how non-linear thinking is inaccurately represented with mainstream chatbot interfaces. For example, when reflecting on design practice in the exercise illustrated in Figure 3, Author B noted "What's interesting in this branching is that I reflected-in-action (#3) on ChatGPTree's response (#2) but after reading ChatGPTree's responses (#4), I branched off to question the context that I started with (#1 and #2), and reflected on the 'purpose' (#5 and #6), then returned to the original conversation (#7 and #8)". This switching between two dialogues suggests that the branching structure supported a reflective cycling between the specifics of the task and its underlying rationale, and that a richer understanding of the context enabled more insightful questions about practice.

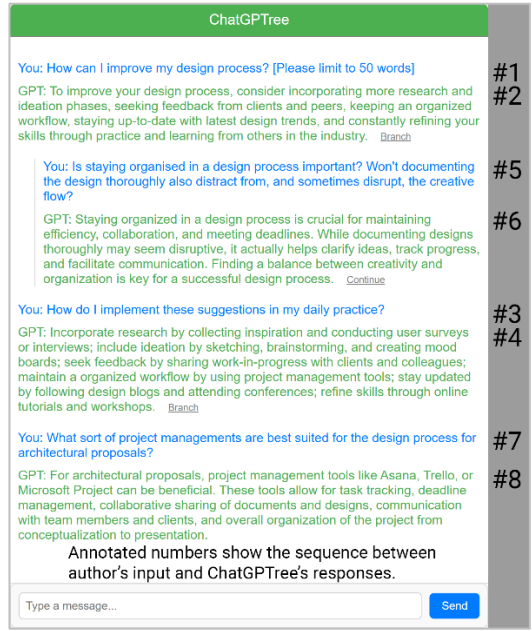

Figure 3 An illustration of a reflective practice exercise in ChatGPTree. For readability of the figure, the LLM has been instructed to keep responses short.

The investigation demonstrated the value of changing the structure of linear LLM dialogue. It allowed users to reflect-in-action more deeply by exploring tangent reflections simultaneously, though these remained visually presented in a linear, chronologically-ordered structure. This format makes it difficult for a user to appreciate the overall shape of their conversation, particularly in longer reflective exercises, and left the experience feeling, in Author A's reflection, "procedural". Additionally, anyone reviewing the conversation would encounter it in this top-to-bottom format, making it harder to discern the timing and connections of the user's reflective moves. This core limitation of providing a logical but not yet visual structure for reflection motivated the development of our next prototype.

### 4.2 ChatBraid: Visualising Reflective Paths for Reflection-on-Action

Building on ChatGPTree, ChatBraid transforms the conversation tree into a node-link diagram (Figure 4). Nodes can be repositioned by the user individually or as part of a sub-graph on a pannable and zoomable canvas, creating an effectively infinite conversational space. By making the entire conversational structure visible and manipulable, the design aimed to encourage users to reflect on how their inquiry could have unfolded differently.

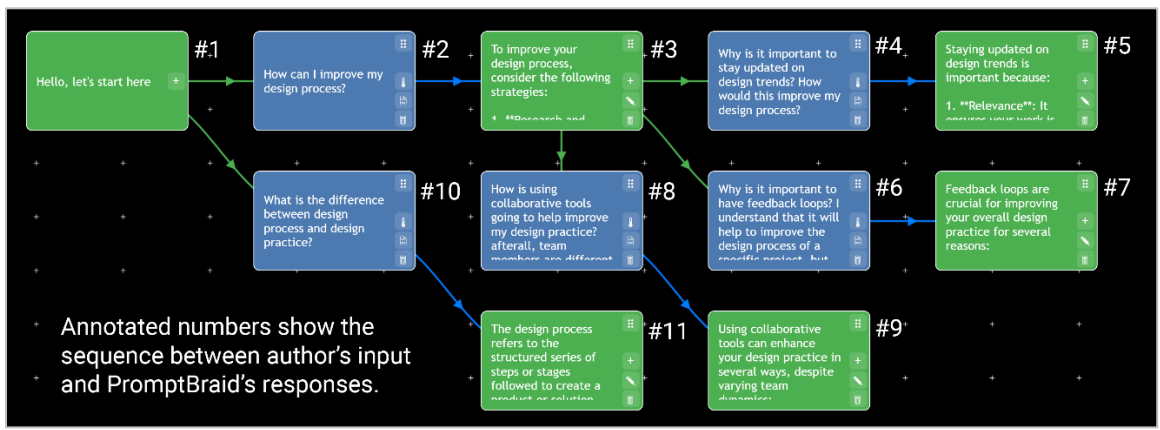

Figure 4 A screenshot of ChatBraid. One interaction from our author-based enquiry is shown.

The visual nature of the interface made it easier to see and reflect on past paths. After iteratively conversing with ChatBraid (as illustrated in Figure 4, #1 to #9), the initial prompt (Figure 4, #2) remained visible on the interface. This contrasts with mainstream LLMs, where the earlier prompts typically scroll off-screen and are easily obscured and to some extent, making us, in effect, forget what we discussed with the chatbot. Keeping the first prompt in view helps users remain aware of how it has shaped their existing conversation, and allows users to easily compare and reflect on the dialogue as a whole (Figure 4, #2 through #9) instead of experiencing it as an ask-and-respond linear exchange (Figure 4, #2 → #3 → #4 etc). By displaying the full conversation, users can revisit earlier messages (Figure 4, #2 to #9), trace the evolution of their thinking, evaluate key moments (Figure 4, #2, #4, #6, and #8) and reflect-on-action (Figure 4, #10 and #11). Although using the same design practice context as for ChatGPTree, we also noted that reflections could become more extensive: we found that the visual branching mechanism encouraged more lines of reflective enquiry, as seen in the branching from node #3 to #4, #6, and #8 in Figure 4.

In ChatBraid, reflection was only implicitly encouraged and relied on users to identify reflective opportunities themselves. Furthermore, the graph structure allowed only one root node, thus limiting users to one conversation on the canvas. These constraints motivated the development of a third prototype that could move beyond this passive, single-conversation structure to provide more active and expansive support for reflection.

### 4.3 ChatGraPhT: Active Scaffolding of Reflections

To address the limitations identified in ChatBraid, we developed the third prototype as a reflective partner interface that integrates visual structure and interaction with agentic LLM guidance. ChatGraPhT (see Figure 5), introduces a novel approach for scaffolding a variety of reflection-in-actions and reflection-on-actions within a chatbot conversation. We extended ChatBraid in two key ways.

First, the graph structure was enhanced to allow users to create and combine multiple reflective threads on a single canvas. In Figure 5, a user has initiated two separate graphs (black #1 and #2) and, later, has merged these by feeding two blue LLM response nodes into a single green user prompt node (black #3). A user prompt node can take any number of parent LLM response nodes. This merging function enables users to synthesise divergent reflections into a cohesive dialogue, making their reflection-on-action more visible – that is, enabling users to step back to integrate insights across different conversation threads.

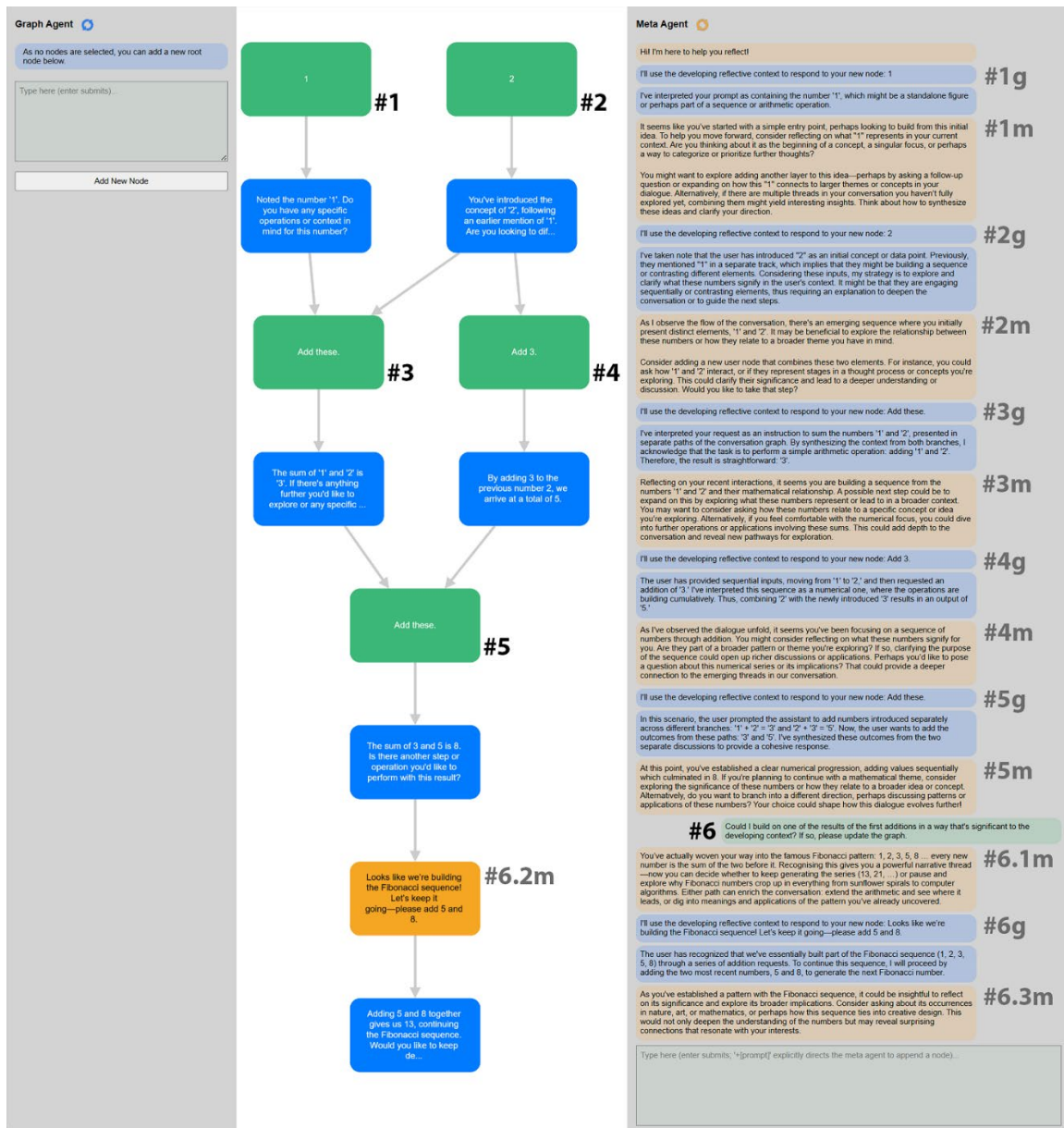

Figure 5 A screen capture of the full ChatGraPhT interface showing the Graph Agent on the left and the Meta Agent on the right. As with Figure 2, a simple mathematical addition exercise is shown. The conversation graph demonstrates merging and branching. Number annotations in black are user actions. Corresponding annotations in grey are following agent actions with the suffixes 'g' for Graph Agent and 'm' for Meta Agent.

Second, ChatGraPhT embeds an agentic framework to provide active reflective guidance. ChatGraPhT comprises two collaborating agents: a Graph Agent and a Meta Agent.

The Graph Agent (Figure 5, left grey sidebar) is where the user converses with the system to build the conversation graph in real time. Rather than prompting the user to reflect explicitly, it supports reflection-in-action by capturing the conversation as a live visual record that users can expand or revisit. When a user selects one or more nodes, it offers options to build from, merge, or edit them as appropriate. If none is selected, it provides the option to add a new root node (as seen in Figure 5 top left). The full text of any single selected node is displayed in its main text area. Its generation of LLM responses to user prompts draws on high-level guidance from the Meta Agent (explained below). To maintain conversational context, the agent can access earlier parts of the conversation, including combined branches, to enable users to carry their reflections across different conversation threads. It also automatically updates all downstream parts of the graph when a user edits a previous message, presenting the conversation as a dynamic network. The Graph Agent may also trigger structural changes, such as generating multiple responses from one prompt to provide different perspectives.

The Meta Agent (Figure 5, right grey sidebar) operates at a higher level, acting as the user's reflection-on-action partner. It monitors the full conversation canvas for patterns, themes, or missed opportunities that could broaden a user's understanding of the conversation. When appropriate, it offers direct suggestions to encourage the user to step back and reflect. The Meta Agent can also directly insert reflective prompt nodes into the graph to guide the user's thinking. These behaviours are illustrated in action in the toy example of Figure 5 in which a user has built a mathematical addition exercise using node branching and merging (black #1 to #5). After their final move (black #5), they see the Meta Agent's suggestion on deeper exploration (grey #5m) and ask it for further advice (black #6). The Meta Agent uncovers the user has unwittingly built the Fibonacci pattern (grey #6.1m) and intervenes in the graph to extend the pattern (grey #6.2m).

Together, these two agents form an interconnected system that links moment-to-moment thinking within a reflective loop. We diagrammed this system in Figure 6. The Meta Agent shares reflection-on-action insights with the Graph Agent, which shapes how the Graph Agent responds to users' prompts. In turn, when the user edits the node-link diagram, the Graph Agent updates the Meta Agent with its reflection-in-action interpretation (Figure 5, #1g to #6g). This two-way exchange creates a feedback loop that supports both the doing (reflection-in-action) and reflecting (reflection-on-action), helping users think through their ideas while also stepping back to see the bigger picture.

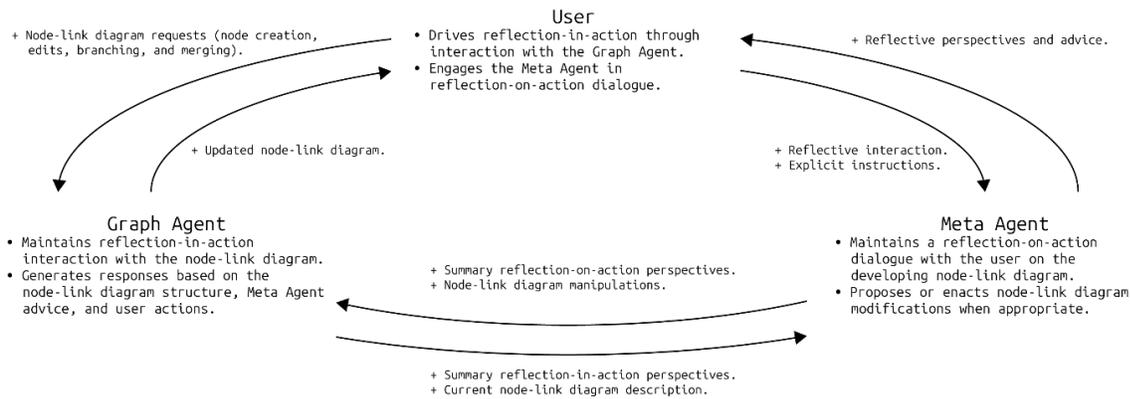

Figure 6 The user–agent and agent–agent interactions and feedback loop in the ChatGraPhT prototype.

We evaluated ChatGraPhT through a first-person enquiry using the same reflective design scenario as in earlier prototypes: "How can I improve my design practice?" From the outset, the system offered distinct and complementary forms of guidance. In response to the initial query (Figure 7, #1), the Graph Agent flagged it would "interpret your inquiry as a general request… without additional context" (Figure 7, #2), before updating the node-link diagram accordingly (Figure 7, #3).  Seeing this interpretation visualised in the node-link diagram prompted immediate reflection in Author A: "I'd been too general, I meant to reflect on my own practice but hadn't actually said that." This moment prompted Author A to reflect on how the enquiry itself had been framed, a meta-level turn that reshaped subsequent interaction.

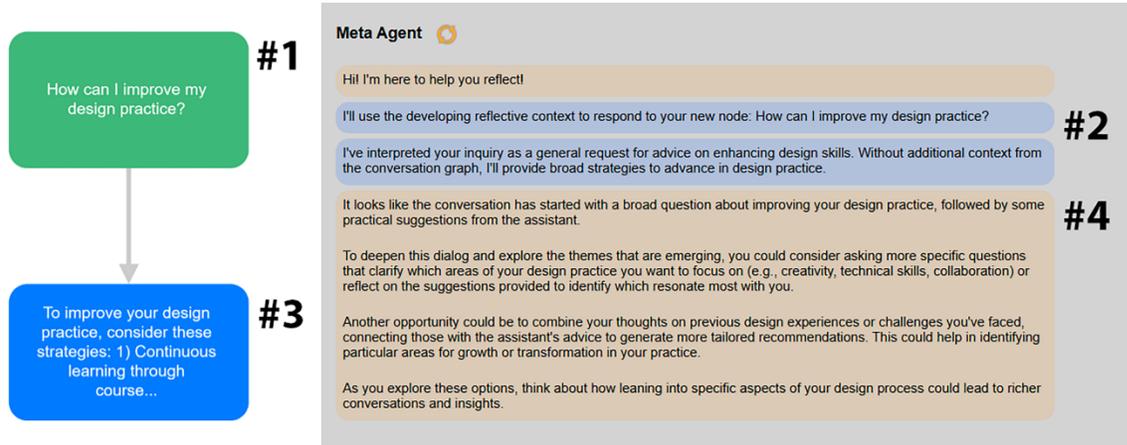

Figure 7 An initial exchange with ChatGraPhT. The root query (#1) was identified by the Graph Agent (#2) as general in scope.

Shortly after, the Meta Agent added a broader intervention (Figure 7, #4), suggesting that the user connect "thoughts on previous design experiences or challenges you've faced" with the Graph  Agent's advice. This reframing invited a shift in stance, prompting a move from exploratory, moment-to-moment thinking toward a more deliberate review of patterns across practice. What began as a basic reflective exchange was elevated to reflection-on-action, with the system offering layered, timely support across both interpretive and metacognitive levels. In response to this shift, the initial prompt was updated in-place to clarify the focus of the reflection (Figure 8).

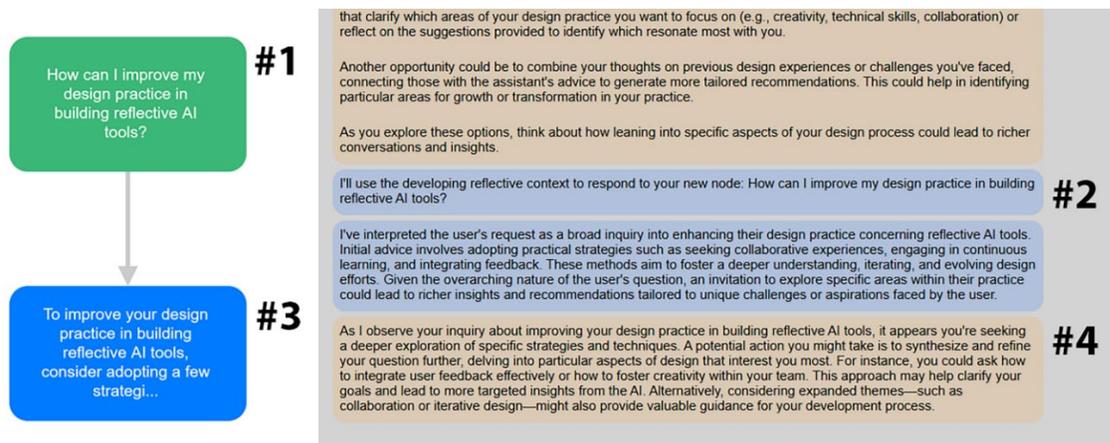

Figure 8 Editing the initial prompt to be more specific in ChatGraPhT (#1), causing an update of the conversation graph (#3; cf. Figure 7). The scheme for numbering of actions matches that of Figure 8.

The updated enquiry, "How can I improve my design practice in building reflective AI tools?", yielded an initial response highlighting feedback loops, transparency, and introspection. Author A pursued two distinct reflective concerns drawn from prior experience: the challenge of feedback being too indirect for users to feel agency (Figure 9, #1) and the difficulty of implementing transparency without overwhelming users (Figure 9, #2). These reflections evolved in parallel, supported by separate branches of the conversation.

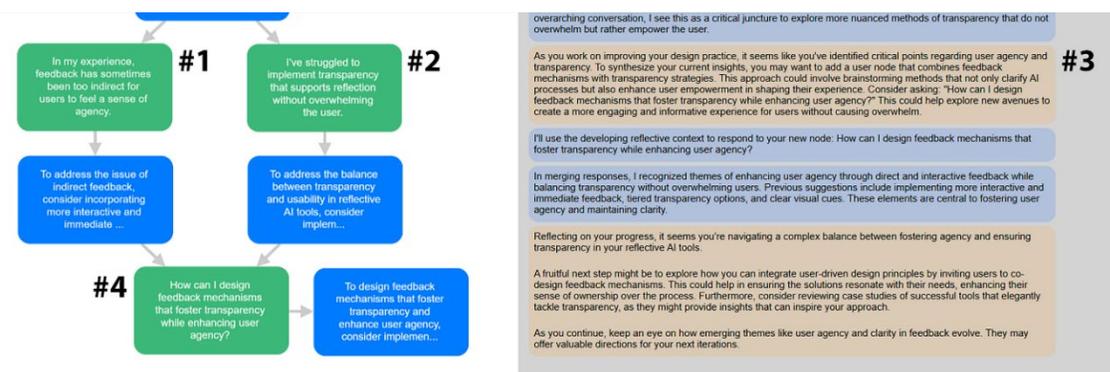

Figure 9 Developing two reflective paths in ChatGraPhT: feedback being too indirect for users to feel agency (#1) and implementing transparency without overwhelming users (#2). The Meta Agent suggested a synthesis (#3), which the user undertook by merging the paths with the suggested prompt (#4).

Shortly after, the Meta Agent offered a higher-level synthesis (Figure 9, #3), suggesting insertion of the question: "How can I design feedback mechanisms that foster transparency while enhancing user agency?" This suggestion was added as a new user prompt to merge both prior paths into a coherent line of inquiry (Figure 9, #4). This intervention prompted a second reflective reframing. Supported by the system's structure and agentic guidance, it marked a shift from parallel reflection-in-action to a moment of reflection-on-action as the user stepped back to synthesise across threads and reframe the inquiry around a broader, more strategic design challenge. While the evaluation described here shows a productive instance, other use suggested that the agentic interventions were sometimes more frequent or generic than needed.

The combination of features in ChatGraPhT transforms it into an active partner in user reflection. By allowing users to branch, edit, and merge their conversations, and offering both direct feedback (Graph Agent) and broader guidance (Meta Agent), our prototype allows users to see the structure of their thinking across multiple lines of inquiry. Users are supported in making sense of their conversations, identifying patterns, comparing alternatives, and extracting insights to guide their future actions. Reflection is no longer left entirely to the users' own initiative.

## 5 DISCUSSION

In this discussion, we situate the early contributions of the ChatGraPhT working prototype's structural and agentic affordances in relation to prior work on node-link LLM interfaces and shared representation research. We then reflect on limitations and future directions.

### 5.1 Integrating Structural and Agentic Scaffolding for Reflective Dialogue

ChatGraPhT responds to two key limitations in current interfaces: the constraint of linear interaction structures [22,31:135–136] and the fragmentation of reflection support that requires external processes [see 3,34]. To address these, it combines a node-link structure with agentic scaffolding to support reflective dialogue. Unlike prior systems [e.g., 2,9,22], which visualise branching using node-link diagrams with a single root and no structural merging, ChatGraPhT treats the conversation as a unified, manipulable graph that allows multiple roots and true merging of branches and separate graphs. This enables users to explore divergent reasoning paths and to also undertake the key capacity for reflection-on-action of synthesis in which two or more conversational paths can come together.

The conversation graph is a persistent shared representation that enables coordinated, bidirectional manipulation by both user and agents. It extends prior node-link systems with limited forms of shared representation [13,14,21,22]. This control enables the Meta Agent to intervene strategically within a parallel conversation channel, such as by offering summarisation, proposing reframing, or suggesting connection to earlier paths, and to operate directly on the node-link diagram. These combined affordances support a shift from passive prompt–response interaction to a co-reflective system that actively scaffolds the user's reasoning process.

### 5.2 Design Reflections on Reflective AI Interfaces

Through a constructive design process involving three prototypes and first-person engagement with each, this research offers generalisable insights into how structural and agentic features shape reflective engagement with LLMs. Persistent visualisation and branching supported reflection-in-action by allowing users to explore alternative paths without obscuring earlier reflection. The ability to revisit earlier branches and merge them enabled reflection-on-action and encouraged re-entry into prior states and synthesis across diverging lines of dialogue. Agentic interventions, such as reflective advice or direct action on the node-link diagram, supported metacognitive shifts by prompting users to reframe or deepen their reflection.

In contrast to approaches in which reflection is scheduled at fixed points in the process [1,36,37], ChatGraPhT embedded reflective opportunities within a live flow of dialogue. This allowed reflection to arise from a user's ongoing engagement, thereby gaining the advantages of a structured approach without its rigid scheduling.

These findings identify both affordances and constraints. Although node-link diagramming and agentic support enhanced the flexibility and depth of reflection, large graphs could become unwieldy and overly frequent interventions risked disrupting reflective flow. Their value appeared greatest when interventions were well-judged and contextually appropriate, pointing to the need for restraint so that agentic support complements rather than displaces reflection. These trade-offs underscore a core tension for reflective AI systems to balance exploratory freedom with cognitive manageability, thus offering support that adapts to user needs without dictating them.

This work extends on Schön's concepts in Reflective Practice [40,41] and of the reflective conversation [42], by enabling both situated reflection-in-action and retrospective synthesis of reflection-on-action within a system that responds to user reasoning structurally and temporally. To support reflective engagement in open-ended tasks, future LLM interfaces could coordinate conversational structure and agent behaviour to offer representations that can be revisited and restructured, alongside interventions that guide metacognitive shifts.

### 5.3 Limitations and Future Work

The visual complexity of large conversational graphs and frequent Meta Agent advice can produce cognitive crowding and disrupt users' reflective focus and sense of continuity. This arises from three interrelated issues: the chronology of the reflection sequence is clearly shown in the Meta Agent's dialogue but is not explicitly detailed elsewhere in the interface; the node-link diagram provides visual and logical structure but yields diminishing clarity as conversational complexity increases; and the Meta Agent may produce commentary too frequently or with insufficient relevance, leading to a sense of interruption.

Future work should aim to improve the organisation of the node-link diagram and agent judgement to support smoother and more sustained reflection. Adding temporal markers could help users track the unfolding of reflection over time and link reflection-on-action to reflection-in-action. Visual grouping of related branches [see 20] may help users to view the overall structure of a growing conversation and identify relationships between subparts of a diagram. Enhancing the Meta Agent's awareness of dialogue structure and reflective processes, such as by focusing more selectively on dialogue structure or generating advice only when significant reflective value is anticipated, could increase the relevance and impact of its interventions.

The evaluation in this study relies on a first-person method [see 12,29], which is appropriate for the early-stage, theory-informed construction of a novel system. This approach provides deep insight into the system's capabilities from an expert perspective but does not capture how external users might interpret, misinterpret, or creatively use the interface. Future work will focus on an empirical user study of an iterated ChatGraPhT, employing validated instruments to assess its effectiveness: the Technology-Supported Reflection Inventory (TSRI) for reflective support [5], the System Usability Scale (SUS) for usability [8], and the NASA Task Load Index (TLX) to quantify cognitive impact [18].

### 5.4 Practice Implications

Externalising reasoning in a structured form may help students visualise thinking to identify gaps in understanding and iteratively reframe problems. For educators, this may expose aspects of learners' decision-making and reflection processes that are typically hidden in conventional dialogue. In professional contexts, such tools could support iterative problem-solving by capturing the evolution of reasoning paths and supporting the retrieval and synthesis of earlier insights.

## 6 CONCLUSION

This paper reports on an early-stage research prototype. Combining a visual dialogue structure with agentic features can support reflective practices. As mainstream LLMs present conversations only as linear transcripts that obscure earlier exchanges and make revisiting or reorganising ideas cumbersome, our ChatGraPhT prototype overcomes their limitation and facilitates situated reflection by (1) transforming conversations into a manipulable, branching network and (2) introducing LLM agents to instigate reflection-in-actions and reflection-on-actions more intuitively. These findings highlight the potential of structured, multi-path LLM interfaces to better organise and advance reflective practice.